\documentclass[12pt]{article}
\usepackage{amsmath}
\usepackage{amsfonts}
\usepackage{amssymb}
\usepackage{srcltx}

\def\ptl{\partial}
\def\oi{b_{\infty }}
\def\di{d_{\infty}}
\def\gli{gl_{\infty }}

\def\Oi{ {B_{\infty }}}

\def\Di{D_{\infty}}
\newtheorem{theorem}{Theorem}
\newtheorem{proposition}[theorem]{Proposition}

\newtheorem{example}[theorem]{Example}

\begin{document}
\hfill\today\\
\noindent
{\Large {\bf Polynomial tau-functions of BKP and DKP hierarchies }}
\vskip 9 mm
\begin{center}
\begin{minipage}[t]{70mm}
{\bf Victor Kac}\\
\\
Department of Mathematics,\\
Massachusetts Institute of Technology,\\
Cambridge, Massachusetts 02139, U.S.A\\
e-mail: kac@math.mit.edu\\
\end{minipage}\qquad
\begin{minipage}[t]{70mm}
{\bf Johan van de Leur}\\
\\
Mathematical Institute,\\
Utrecht University,\\
P.O. Box 80010, 3508 TA Utrecht,\\
The Netherlands\\
e-mail: J.W.vandeLeur@uu.nl
\end{minipage}
\end{center}
\begin{abstract}
We construct all polynomial tau-functions of the BKP, DKP and MDKP hierarches.
\end{abstract}
\section{Introduction}
In his seminal paper \cite{S}  M. Sato introduced the KP hierarchy of evolution equations of Lax type 
\begin{equation}
\label{A1} 
\frac{dL}{dt_n} = [ (L^n)_+, L  ], \ n = 1, 2, \ldots, 
\end{equation}
on the pseudodifferential operator $ L = \ptl + u_1(t) \ptl^{-1} + u_2(t) \ptl^{-2} + \ldots$,   where $t=(t_1,t_2,\ldots)$ and
$ \ptl = \frac{\ptl}{\ptl t_1} $.
He introduced the corresponding tau-function $\tau(t)$ and showed that any solution $u_1(t), u_2(t),\ldots$  of (\ref{A1})  can be expressed as a differential polynomial in $\tau$. He also showed that the tau-functions  form an infinite Grassmann manifold of type $A$, and that the set of polynomial tau-functions includes all Schur polynomials $s_\lambda(t)$, where $\lambda\in \rm  Par$, the set of partitions. His ideas have been subsequently developed by his school in the series of papers,
including \cite{DJKM1},  \cite{DJKM2},  \cite{DJKM3},  \cite{JM}.

The totality of polynomial tau-functions of the Kp hierarchy is a disjoint union of Schubert cells $\coprod_{\lambda\in\rm Par}C_\lambda$, such that $s_\lambda(t)$ is the "center" of $C_\lambda$. In our paper \cite{KvdLmodKP} we proved, using the boson-fermion correspondence, that each $C_\lambda$ can be obtained from $s_\lambda(t)$ by shifts of  the $t_i$ by certain constants.

In the papers \cite{DJKM3}, \cite{JM} the BKP and DKP hierarchies have been introduced along the lines proposed by Sato for KP. In his papers \cite{You} and \cite{You2}
Y. You proved that the $Q$-Schur polynomials are polynomial tau-functions  of the BKP, DKP and MDKP hierarchies. As in the KP case, these tau-functions are "centers" of Schubert cells of the infinite-dimensional orthogonal Grassmann manifold. In the present paper we construct all polynomial tau-functions for the BKP, DKP and MDKP hierarchies in each of the Schubert cells, using the boson-fermion correspondence of types $B$ and $D$ \cite{KvdLB}. 

Note that, as shown in \cite{A}, the CKP hierarchy has no polynomial tau-functions.

\section{The spin representation of $\oi$ and $\di$
and the BKP and DKP hierarchy in the fermionic picture}
Consider the Lie algebra over $\mathbb{C}$,
$$\gli =\{(a_{ij})_{i,j\in {\bf Z}}|  \ \hbox{ all but a finite number of }a_{ij}\in\mathbb{C}\  \hbox{are zero} \}.$$
The matrices $E_{ij}$ with  $(i,j)$--th entry 1 and 0 elsewhere,
for $i,j\in{\bf Z}$ form a basis.
Define on $\gli$ the following two linear anti--involutions:
$$\iota_B(E_{jk})=(-1)^{j+k}E_{-k,-j}\quad 
\iota_D(E_{jk})=E_{-k+1,-j+1}. $$
Using these anti--involution we define the Lie algebras $\oi$  and $d_\infty$ as
a subalgebra of $\gli$:
$$\oi=\{a\in\gli |\iota_B(a)=-a\},\quad \di=\{a\in\gli |\iota_D(a)=-a\}.$$
The elements $F_{jk}=E_{-j,k}-(-1)^{j+k}E_{-k,j}=-(-1)^{j+k}F_{kj}$, with $j>k$, respectively 
$G_{j+\frac12, k-\frac12}=E_{-j,k}-E_{-k+1,j+1}=- G_{k-\frac12, j+\frac12} $, with $j\ge k$,
 form a basis of $\oi$, respectively $\di$.
We have the following root space decomposition of $\oi$:
\[
\oi= \frak{h}\oplus\bigoplus_{\alpha\in \Delta^B_+} \left(\frak{g}_{-\alpha}\oplus\frak{g}_{\alpha}\right),\ \mbox{where }\frak{h}=\bigoplus_{i>0}\mathbb{C}F_{-i,i}\, ,
\] 
\[
\Delta^B_+=\{ \epsilon_i|\, i=1,2,\ldots\}\cup \{ \epsilon_i\pm \epsilon_j|\, i>j>0\},\quad
\frak{g}_{\pm \epsilon_i}=\mathbb{C}F_{\pm i,0}, \ \frak{g}_{\epsilon_i\pm \epsilon_j}=\mathbb{C}F_{i,\pm j}\, .
\]
The  root space decomposition of $\di$ is as follows:
\[
\di= \frak{h}\oplus\bigoplus_{\alpha\in \Delta^D_+} \left(\frak{g}_{-\alpha}\oplus\frak{g}_{\alpha}\right),\ \mbox{where }\frak{h}=\bigoplus_{i>0}\mathbb{C}G_{-i,i}\, ,
\] 
\[
\Delta^D_+=\{ \epsilon_i\pm \epsilon_j|\, i>j\ge 0\},\quad
 \frak{g}_{\epsilon_i\pm \epsilon_j}=\mathbb{C}G_{i+\frac12, \pm (j+\frac12)} .
\]
We now describe the spin representation of $\oi$ and $\di$.  For this purpose we introduce the Clifford algebras $BC\ell$  and $DC\ell$ as the
associative algebras generated by the vector space $\mathbb{C}^\infty$ with basis $\phi_j$, $j\in{\bf Z}$, respectively $j\in\frac12 +{\bf Z}$,   and symmetric bilinear forms 
\begin{equation}
\label{1}
(\phi_i,\phi_j)_B=(-1)^i\delta_{i,-j},\quad (\phi_i,\phi_j)_D=\delta_{i,-j}
\end{equation}
 with defining
relations
\begin{equation}
\label{1.1} vw+wv=(v,w), \ v,w \in\mathbb{C}^\infty\, .
\end{equation}
We have a $\mathbb{Z} /
2\mathbb{Z}$-gradation
\[
BC\ell\ = BC\ell_{\overline 0}\oplus BC\ell_{\overline{1}},\quad\mbox{resp. }DC\ell\ = DC\ell_{\overline 0}\oplus DC\ell_{\overline{1}}
\]
where $BC\ell_\nu $ $ (\nu \in \mathbb{Z} / 2\mathbb{Z})$ is
spanned by all products of $s$ elements of $\mathbb{C}^\infty$  with $s\equiv \nu$
mod $2$. We shall identify $\mathbb{C}^\infty$ with its image in $BC\ell
$ and $DC\ell$.

We define the spin module $V_B$ over $BC\ell$  and $V_D$ over $DC\ell$ as the irreducible module with
highest weight vector, the {\it vacuum vector} $|0\rangle\ne 0$, satisfying
$$\phi_j|0\rangle=0\quad\hbox{ \rm for}\ j>0, $$
where in the $B$-case we also assume that
\[
\phi_0|0\rangle =\frac1{\sqrt 2}|0\rangle. 
\]
The elements $\phi_{j_1}\phi_{j_2}\cdots\phi_{j_p}|0\rangle$
with $j_1<j_2<\cdots <j_p< 0$ form a basis of $V_B$ (here all $j_k\in\mathbb{Z}$) and $V_D$ (here all $ j_k\in \frac12 +\mathbb{Z}$). Then we obtain the representation of $\oi$, respectively $\di$, by 
$$\pi_B(F_{jk})={(-1)^j\over 2}(\phi_j\phi_k-\phi_k\phi_j),\quad\mbox{resp. }\pi_D(G_{jk})=\frac12 (\phi_j\phi_k-\phi_k\phi_j)\, .
$$
It is irreducible in the $B$-case. If we restrict to $\di$, the module  splits into two irreducible 
$\di$-modules $V_D=V_0\oplus V_1$, with highest weight vectors $|0\rangle$ and $|1\rangle=\phi_{-\frac12}|0\rangle$.

{\it From now on let $G=B$ or $D$, we will write e.g. $GC\ell$ for the corresponding Clifford algebra.}

It will be convenient also to introduce the opposite $\oi$ and $\di$-module with highest weight vector  $\langle 0|$, by
\[
\langle 0|\phi_0=\frac1{\sqrt 2}\langle 0|,\quad
\langle 0|\phi_j=0,\ \mbox{for }j<0.
\]
For $\di$ we also have $\langle 1|=\langle 0|\phi_{\frac12}$.
The 
{\it vacuum expectation value of} $a\in GC\ell$   is defined as $\langle 0|a|0\rangle\in\mathbb{C}$.

Given a non-isotropic vector $\alpha\in\mathbb{C}^\infty$ (i.e. $(\alpha ,
\alpha) \neq 0)$, the associated reflection $r_{\alpha}$ is defined by
\[
r_{\alpha} (v) = v - \frac{2(\alpha, v)}{(\alpha, \alpha)} \alpha .
\]
Let $GC\ell^\times$ denote the multiplicative group of invertible
elements of the algebra $GC\ell$. We denote by ${\rm Pin}_G \mathbb{C}^\infty$ the subgroup
of $GC\ell^\times$ generated by all the elements $a$ such that
$a\mathbb{C}^\infty a^{-1}=\mathbb{C}^\infty$ and let ${\rm Spin}_G \mathbb{C}^\infty={\rm Pin }_G\mathbb{C}^\infty\cap GC\ell_{\overline{0}}$.

If $\alpha \in \mathbb{C}^\infty$ is a non-isotropic vector, then, by (\ref{1.1}):
\begin{equation}
  \label{1.3}
  \alpha^{-1} = \frac{2\alpha}{(\alpha , \alpha)} ,
\end{equation}
hence $\alpha \in GC\ell^\times$, and from (\ref{1.1})  we
obtain
\begin{equation}
  \label{1.4}
  \alpha v \alpha^{-1} = -r_{\alpha} (v) ,
\end{equation}
hence $\alpha \in \;\mbox{Pin}_G\mathbb{C}^\infty$. We have a homomorphism $T:\,{\rm  Pin}_G
\mathbb{C}^\infty \rightarrow \Oi $ or $\Di,\  g\mapsto T_g$ defined by $(v\in \mathbb{C}^\infty)$:
\[
T_g (v)= g v g^{-1}  \in \mathbb{C}^\infty.
\]
Here $\Oi$ or $\Di$ is the subgroup of  the group
$$GL_{\infty}= \{ (g_{ij})_{i,j \in \mathbb{Z}}\ \mbox{which are 
 invertible and all but a finite number of }g_{ij} -
\delta_{ij}\ \mbox{ are }0\},$$
consisting of elements which preserve the bilinear form (\ref{1}). Thus, we have a projective representation of $\Oi$ on $V_B$ and $\Di$ on $V_D$.

The orthogonal $G$-Grassmannian is the collection of all linear subspaces 
\[
	{\rm Ann}_G f = \{ v\in {\mathbb{C}^\infty} \mid v f = 0\},\quad \mbox{for all f}\in O.
\]
Each ${\rm Ann}_G f $ is a maximal isotropic subspace of $\mathbb{C}^\infty$. In fact,  the orthogonal Grassmannian is the collection of all maximal isotropic subspaces $U$ of $\mathbb{C}^\infty$ such that $\phi_{j}\in U$ for all $j>>0$.

Let us focus on the $B$-case first, which is well-known, see e.g. \cite{DJKM3} or \cite{You}.
The group $\Oi$ is the subgroup of  $GL_\infty $ generated by all $-r_{\alpha}$. 
Let  ${\cal O}=\left({\rm Spin}_B\mathbb{C}^\infty\right)|0\rangle$. Since $\phi_0\in{\rm Pin}_B\mathbb{C}^\infty$ and $\phi_0|0\rangle=\frac1{\sqrt 2}|0\rangle$, one has that ${\cal O}  ={\rm Pin}_B\mathbb{C}^\infty|0\rangle$.

Let $S_B$  be the following operator on $V_B\otimes V_B$:
$$
\label{S}
S_B=\sum_{j\in\mathbb{Z}} (-1)^j \phi_j\otimes \phi_{-j},$$
Then (see e.g. \cite{KvdLB})
\begin{theorem}
\label{t1.8}
If $\tau \in V_B$ and $\tau\ne 0$, then $\tau \in {\cal O}$ if only if
$\tau$
    satisfies the equation
    \begin{equation}
      \label{1.9}
      S_B(\tau \otimes \tau) = \frac{1}{2} \tau \otimes \tau.
    \end{equation}
\end{theorem}
Equation (\ref{1.9})  is called the fermionic BKP 
hierarchy.

In the $D$-case, let  ${\cal O}_0=\left({\rm Spin}_D\mathbb{C}^\infty\right)|0\rangle$ and ${\cal O}_1=\left({\rm Spin}_D\mathbb{C}^\infty\right)|1\rangle$. In this case, let 
$$
\label{SD}
S_D=\sum_{j\in\frac12 + \mathbb{Z}}  \phi_j\otimes \phi_{-j},$$
Then (see e.g. \cite{You2} or \cite{KvdLB})
\begin{theorem}
\label{t1.9}
(a)  If $\tau \in V_\nu$ and $\tau\ne 0$, then $\tau \in {\cal O}_\nu$ if only if
$\tau$
    satisfies the equation
    \begin{equation}
      \label{1.9D}
      S_D(\tau \otimes \tau) =0
    \end{equation}
(b) A pair of non-zero  DKP tau-functions $\tau_0\in {\cal O}_0$ and $\tau_1\in {\cal O}_1$ satisfies the modified DKP hierarchy, i.e.,
\begin{equation}
\label{modDKP}
 S_D(\tau_0 \otimes \tau_1) =  \tau_1 \otimes \tau_0.
    \end{equation}
if and only if
\[
{\rm dim} \left(({\rm Ann} \;  \tau_0+{\rm Ann} \;  \tau_1)\slash ({\rm Ann} \;  \tau_0\cap {\rm Ann} \;  \tau_1)\right)=2\, .
\]
\end{theorem}

If $\tau$ is a BKP (resp. DKP) tau-function and  $\alpha\in \mathbb{C}^\infty$, then it is shown in \cite{KvdLB}, Lemmas  2.1 and 2.2, that $\alpha \tau$ is again a BKP (resp. DKP) tau-function. Hence, since $|0\rangle$ is a tau-function,
\begin{equation}
\label{orbit}
\begin{aligned}
v_1v_2\cdots v_k|0\rangle\in {\cal O}\cup \{0\}\quad\mbox{for any }v_i\in\mathbb{C}^\infty \, ,\\
v_1v_2\cdots v_{2\ell}|0\rangle\in {\cal O}_0\cup \{0\}\quad\mbox{for any }v_i\in \mathbb{C}^\infty \, ,\\
v_1v_2\cdots v_{2\ell+1}|0\rangle\in {\cal O}_1\cup \{0\}\quad\mbox{for any }v_i\in \mathbb{C}^\infty \, .
\end{aligned}
\end{equation}
Without loss of generality we may choose, in every $v_j$, the coefficient of the lowest $\phi_n$ equal to 1, in other words we assume that all $v_j$ are of the form
\begin{equation}
\label{orbit1}
\begin{aligned}
B:&\qquad
v_j=\phi_{-\lambda_j}+\sum_{n<\lambda_j} a_{-n,j}\phi_{-n},\ \mbox{with }
a_{-nj}=0\ \mbox{for }n<<0,\\
D:&\qquad
v_j=\phi_{-\lambda_j- \frac12}+\sum_{n<\lambda_j+\frac12 } a_{-n,j}\phi_{-n},\ \mbox{with }
a_{-nj}=0\ \mbox{for }n<<0,
\end{aligned}
\end{equation}
where the $\lambda_j$ are positive integers in the $B$-case and non-negative integers for $D$. 
We may also assume that $\lambda_1>\lambda_2>\cdots >\lambda_k$. 

\section{Schubert cell decomposition of the orbit}

For $G=B$, let  $W_0$ be the subgroup of the Weyl group  generated by reflections in the short roots $\epsilon_i$, i.e. $w\in W_0$ if and only if $w(\epsilon_j)=\pm \epsilon_j$, where only finitely many $\epsilon_j$ are mappped to $-\epsilon_j$. 
There is a one to one correspondence between the set of all strict partitions and the elements of $W_0$.
Namely, we associate to a strict partition
 $\lambda=(\lambda_1,\lambda_2,\ldots,\lambda_k)$, i.e., $\lambda_i>\lambda_{i+1}$   the following element of  $W_0$:  $w_\lambda (\epsilon_{\lambda_n})=-\epsilon_{\lambda_n}$ for $n=1,2,\ldots,k$ and all other $\epsilon_j$ fixed. 

For $G=D$, let  $W_0$ be the subgroup of the Weyl group consisting of elements  $w$ such that  $w(\epsilon_j)=\pm \epsilon_j$, now $j\ge 0$ where only a finite even number of  $\epsilon_j$ are mappped to $-\epsilon_j$. The correspondence with strict partitions as in the $B$-case is almost the same,   now  we have an even number of parts, but the last one can be zero, i.e., $\lambda_1>\lambda_2>\cdots  >\lambda_{2k}\ge 0$ corresponds to  $w_\lambda (\epsilon_{\lambda_n})=-\epsilon_{\lambda_n}$ $n=1,2,\ldots, 2k$ and all other $\epsilon_j$ fixed. 
If we allow that a part is also 0, we call such a partition an extended strict partition.

Let $ P_B$ be the parabolic subgroup that corresponds to all long simple roots $\epsilon_{i+1}-\epsilon_i$, for $i=1,2,\dots$, i.e., $P_B$ corresponds to the roots $\epsilon_i-\epsilon_j$, $\epsilon_i+\epsilon_j$ and $\epsilon_i$ for $i\ne j> 0$, thus the corresponding root vectors annihilate the vacuum.
Then we have the Bruhat decomposition
\begin{equation}
\label{bruhat}
\Oi=\bigcup_{w\in W_0}U_w w P_B \quad\mbox{disjoint union},
\end{equation}
where $U_{w_\lambda}$ consists of all elements $\exp \left(\sum_{\alpha} t_\alpha e_\alpha\right)$, where $e_\alpha$ is a root vector corresponding to $\alpha\in \Delta_+$ such that $w_\lambda(-\alpha)\in\Delta_+$, and $t_\alpha\in\mathbb{C}$.

For $G=D$ we need two decompositions corresponding to a different parabolic subgroups, viz $P_D^0$ and $P_D^1$, where both correspond to the simple roots $\epsilon_{i+1}-\epsilon_i$ for $i\ge $ together with  $\epsilon_1-\epsilon_0$ for $P_D^0$, respectively  $\epsilon_1+\epsilon_0$ for  $P_D^1$. Hence, these parabolics correspond to the root vectors $\epsilon_i-\epsilon_j$ and $\epsilon_i+\epsilon_j$, where $i\ne j\ge 0$ for $P_D^0$, respectively $\epsilon_i-\epsilon_j$, where $i\ne j> 0$, $\epsilon_i+\epsilon_j$, where $i>j\ge 0$ and $\pm \epsilon_j-\epsilon_0$ for $P_D^1$.
We have the Bruhat decomposition
\begin{equation}
\label{bruhatD}
\Di=\bigcup_{w\in W_0}U_w w P_D^i \quad\mbox{disjoint union},\ i=0\ \mbox{or }1,
\end{equation}
where $U_{w_\lambda}$ consists of all elements $\exp \left(\sum_{\alpha} t_\alpha e_\alpha\right)$, where $e_\alpha$ is a root vector corresponding to $\alpha\in \Delta_+$ such that $w_\lambda(-\alpha)\in\Delta_+$, and $t_\alpha\in\mathbb{C}$.

Using the Bruhat decompositions (\ref{bruhat}), respectively  (\ref{bruhatD}), and that $P_G|0\rangle=\mathbb{C^\times}|0\rangle$, respectively  
$P_G^i|i\rangle=\mathbb{C^\times}|i\rangle$,
we obtain the Schubert cell decomposition of ${\cal O}$, ${\cal O}_0$ and ${\cal O}_1$:
\begin{equation}
\label {schubert}
\begin{aligned}
{\cal O}&=\bigcup_{w_\lambda\in W_0} U_{w_\lambda}w_\lambda|0\rangle\quad\mbox{disjoint union}, \\
{\cal O}_i&=\bigcup_{w_\lambda\in W_0} U_{w_\lambda}w_\lambda|i\rangle\quad\mbox{disjoint union, }i=0\ \mbox{or }1.
\end{aligned}
\end{equation}
One has  (see e.g. \cite{You}):
\[
w_\lambda|0\rangle=|\lambda \rangle=\phi_{-\lambda_1}\phi_{-\lambda_2}\cdots
\phi_{-\lambda_k}|0\rangle\, \ \mbox{for } G=B
\] 
and for $G=D$:
\[
\begin{aligned}
w_\lambda|0\rangle=&|\lambda \rangle=\phi_{-\lambda_1-\frac12}\phi_{-\lambda_2-\frac12}\cdots
\phi_{-\lambda_{2k}-\frac12}|0\rangle\\
w_\lambda |1\rangle=&
\begin{cases}
|(\lambda_1,\ldots, \lambda_{2k},0)\rangle=\phi_{-\lambda_1-\frac12}\phi_{-\lambda_2-\frac12}\cdots
\phi_{-\lambda_{2k}-\frac12}\phi_{-\frac12}|0\rangle\,&\mbox{if } \lambda_{2k}\ne 0\\
|(\lambda_1,\ldots, \lambda_{2k-1})\rangle=
\phi_{-\lambda_1-\frac12}\phi_{-\lambda_2-\frac12}\cdots
\phi_{-\lambda_{2k-1}-\frac12}|0\rangle\,&\mbox{if } \lambda_{2k}= 0\\
\end{cases}
\end{aligned}
\] 
Hence we have to calculate how an element of  $U_{w_\lambda}$ acts on this vector.
We describe the case $G=B$, the case that $G=D$ can be determined in a similar way.
Note that we should only take vectors $e_\alpha$ such that $e_\alpha w_\lambda|0\rangle$ is non-zero. So  we may remove all positive roots  $\epsilon_j+\epsilon_{\lambda_n}$ where $j<\lambda_n$ and $j\not\in \lambda$.
Hence,  we may assume that an element of $\Oi$ is of the form  $\exp \left(\sum_{\alpha\in\Delta_\lambda} t_\alpha e_\alpha\right)$, where
\[
\Delta_\lambda=\{\epsilon_{\lambda_n}, \epsilon_{\lambda_n}-\epsilon_j, \ \epsilon_{\lambda_n}+\epsilon_{\lambda_m} |
\ 
n,m=1,2,\ldots,k,\ 
0< j<\lambda_n, \ j\not =\lambda_{n+1},\lambda_{n+2},\ldots, \lambda_k \},
\]
We have:
\[
\begin{aligned}
\exp \left(\sum_{\alpha\in\Delta_\lambda} t_\alpha e_\alpha\right)=&
\exp\left(\sum_{1\le m<n\le k} t_{\lambda_m,\lambda_n}
\phi_{\lambda_m} \phi_{\lambda_n}\right)
\exp\left(\sum_{n=1}^k
\sum_{0<j<\lambda_n, j\not\in\lambda} t_{-j,\lambda_n}
\phi_{-j} \phi_{\lambda_n}\right)\times\\
&\qquad\exp\left(\sum_{n=1}^k t_{\lambda_n}
\phi_0 \phi_{\lambda_n}\right),
\end{aligned}
\]
Note that:
\[
\begin{aligned}
\exp\left(\sum_{1\le m<n\le k} t_{\lambda_m,\lambda_n}
\phi_{\lambda_m} \phi_{\lambda_n}\right)=& \prod_{1\le m<n\le k} \left( 1+t_{\lambda_m,\lambda_n}
\phi_{\lambda_m} \phi_{\lambda_n}\right),\\
\exp\left(\sum_{n=1}^k
\sum_{0<j<\lambda_n, j\not\in\lambda} t_{-j,\lambda_n}
\phi_{-j} \phi_{\lambda_n}\right)=&
\prod_{n=1}^k\prod_{0<j<\lambda_n, j\not\in\lambda}\left( 1+ t_{-j,\lambda_n}
\phi_{-j} \phi_{\lambda_n}\right),\\
\exp\left(\sum_{n=1}^k t_{\lambda_n}
\phi_0 \phi_{\lambda_n}\right)=&
1+\sum_{n=1}^kt_{\lambda_n}
\phi_0 \phi_{\lambda_n}\, .
\end{aligned}
\]
Hence applying conjugation several times gives respectively
\[
\left(1+\sum_{n=1}^kt_{\lambda_n}
\phi_0 \phi_{\lambda_n}\right)\phi_{-\lambda_i}\left(1-\sum_{n=1}^kt_{\lambda_n}
\phi_0 \phi_{\lambda_n}\right)=\phi_{-\lambda_i}+(-1)^{\lambda_i}t_{\lambda_i}\phi_0,
\]
\[
\begin{aligned}
\prod_{n=1}^k
\prod_{0<j<\lambda_n, j\not\in\lambda}\left( 1+ t_{-j,\lambda_n}
\phi_{-j} \phi_{\lambda_n}\right)
\left(
\phi_{-\lambda_i}+(-1)^{\lambda_i}t_{\lambda_i}\phi_0
\right)
\prod_{m=1}^k
\prod_{0<j<\lambda_m, j\not\in\lambda}\left( 1- t_{-j,\lambda_m}
\phi_{-j} \phi_{\lambda_m}\right)=\\
\phi_{-\lambda_i}+(-1)^{\lambda_i}\left(t_{\lambda_i}\phi_0+
\sum_{0<j<\lambda_i, j\not\in\lambda} t_{-j,\lambda_i}\phi_{-j}\right),
\end{aligned}
\]
and 
\[
\begin{aligned}
&\exp \left(\sum_{\alpha\in\Delta_\lambda} t_\alpha e_\alpha\right)\phi_{-\lambda_i}\exp \left(-\sum_{\alpha\in\Delta_\lambda} t_\alpha e_\alpha\right)=\\
&\prod_{1\le m<n\le k} \left( 1+t_{\lambda_m,\lambda_n}
\phi_{\lambda_m} \phi_{\lambda_n}\right)\left(
\phi_{-\lambda_i}+(-1)^{\lambda_i}\left(t_{\lambda_i}\phi_0+
\sum_{0<j<\lambda_i, j\not\in\lambda} t_{-j,\lambda_i}\phi_{-j}\right)\right)\times \\
&\qquad\prod_{1\le m<n\le k} \left( 1-t_{\lambda_m,\lambda_n}
\phi_{\lambda_m} \phi_{\lambda_n}\right)\\
&=\phi_{-\lambda_i}+(-1)^{\lambda_i}\left(t_{\lambda_i}\phi_0+
\sum_{0<j<\lambda_i, j\not\in\lambda} t_{-j,\lambda_i}\phi_{-j}+
\sum_{0<\ell<i} t_{\lambda_\ell,\lambda_i}\phi_{\lambda_\ell}-\sum_{i<\ell\le k} t_{\lambda_i,\lambda_\ell}\phi_{\lambda_\ell}
\right)
\, .
\end{aligned}
\]
So for a strict partition $\lambda=(\lambda_1,\lambda_2,\cdots \lambda_k)$ , let
\[
C_{\lambda}= U_{w_\lambda}w_\lambda|0\rangle\, .
\]
From the observation on $U_{w_\lambda}$ above, we find that  the dimension of the corresponding Schubert cell $C_\lambda$ is $|\Delta_\lambda|=|\lambda|$. Moreover, from the calculations above, we find that   the element  (\ref{orbit}) with $v_j$ given by  (\ref{orbit1}) must be an element of $C_\lambda$.
The calculations for $G=D$ can be done in a similar way.
We thus obtain
\begin{proposition}
\label{prop}
(a) Let $SP$ be the set of strict partitions $\lambda=(\lambda_1,\lambda_2,\cdots \lambda_k)$, i.e.,  $\lambda_1>\lambda_2>\cdots>\lambda_k>0$. Then 
\[
{\cal O}=\bigcup_{\lambda\in SP}C_\lambda \qquad \mbox{disjoint union},
\]
where
\begin{equation}
\label{Clambda}
C_\lambda= \{ a\left(\phi_{-\lambda_1}+\sum_{j>-\lambda_1} a_{j1}\phi_{j}\right)\left(\phi_{-\lambda_2}+\sum_{j>-\lambda_2} a_{j2}\phi_{j}\right)\cdots
\left(\phi_{-\lambda_k}+\sum_{j>-\lambda_k} a_{jk}\phi_{j}\right)|0\rangle \}
\end{equation}
and $ a, a_{j\ell}\in\mathbb{C},\ a\ne 0$.
\\
(b)
Let $ESP$ be the set of extended strict partitions $\lambda=(\lambda_1,\lambda_2,\cdots \lambda_k)$, i.e.,  $\lambda_1>\lambda_2>\cdots>\lambda_k\ge 0$, and let $ESP_0$, respectively $ESP_1$,  be the subsets of $ESP$, consisting  of extended strict partitions $\lambda=(\lambda_1,\lambda_2,\cdots \lambda_k)$ with $k$ even, respectively odd. Then 
\[
{\cal O}_i=\bigcup_{\lambda\in ESP_i}D_\lambda \quad (i=0,1),
\]
where
\begin{equation}
\label{Dlambda}
D_\lambda= \{ a\left(\phi_{-\lambda_1-\frac12}+\sum_{j>-\lambda_1} a_{j1}\phi_{j}\right)\left(\phi_{-\lambda_2-\frac12}+\sum_{j>-\lambda_2} a_{j2}\phi_{j}\right)\cdots
\left(\phi_{-\lambda_k-\frac12}+\sum_{j>-\lambda_k} a_{jk}\phi_{j}\right)|0\rangle \}
\end{equation}
and $ a, a_{j\ell}\in\mathbb{C}$,  for all $1\le \ell\le k$ and $ a\ne 0$.

\end{proposition}

\section {Vertex operators and the BKP hierarchy
in the bosonic picture}\
In this section we  will describe the case that $G=B$, the case that $G=D$, will be  done in the next section.

Define the following two generating series:
\[
\phi (z) = \sum\limits_{j\in \mathbb{Z}} \phi_j z^{-j}, \qquad
\alpha(z) = \sum\limits_{k\in 2 \mathbb{Z}+1} \alpha_k
 z^{-k-1} = :
\phi(z)  \frac{\phi(-z)}{z}:\, .
\]
Then one has (see e.g. \cite{KvdLB} for details):
\begin{theorem}(~\cite{DJKM2})
  \label{t2.3}
\[
\phi(z) = \frac{1}{\sqrt{2}} \exp (-\sum\limits_{k<0, \mbox{odd}}
\frac{\alpha_k}{k} z^{-k} ) \exp ( -\sum\limits_{k>0,\mbox{odd}}
\frac{\alpha_k}{k} z^{-k}) .
\]
\end{theorem}

The neutral (twisted) boson-fermion correspondece consists of
identifying the space $V$ with the space $B=\mathbb{C} [ t_1 , t_3 ,
t_5 , \ldots ]$ via the vector space isomorphism
$
\sigma : F \rightarrow B
$
given by
\[
\sigma (\alpha_{-m_1} \alpha_{-m_2} \ldots \alpha_{-m_s} |0\rangle ) = m_1
m_2 \ldots m_s t_{m_1} t_{m_2} \ldots t_{m_s}.
\]
The transported action of the operators $\alpha_m$ is as follows
\[
\begin{array}[h]{lcl}
\sigma \alpha_{-m} \sigma^{-1} (p(t)) &=& m t_m p(t),\\[3mm]
\sigma \alpha_m \sigma^{-1} (p(t)) &=&\displaystyle  2 \frac{\partial
p(t)}{\partial t_m}.
\end{array}
\]
Then
\[
\sigma \phi(z) \sigma^{-1} =\frac{1}{2} \sqrt{2} e^{\xi (t,z)} e^{-\eta
(t,z)},\quad \mbox{where}
\]

\[
\xi (t,z) = \sum\limits_{i=0}^{\infty} t_{2i + 1} z^{2i + 1}, \quad
\eta (t,z) = \sum\limits^{\infty}_{i=0} \frac{2}{2i+1}
\frac{\partial}{\partial t_{2i + 1}} z^{-2i -1}.
\]

We now  rewrite the BKP hierarchy (\ref{1.9}),
using ${\rm Res}_{z=0} dz \sum_j f_jz^j=f_{-1}$. Namely, by Theorem \ref{t1.8}, we have that  $0\ne \tau(t)\in B$ is an element of $\sigma({\cal O})$   if and only if 
\begin{equation}
\label{BKPB}
\begin{aligned}
{\rm Res}_{z=0}{dz\over z}&
\exp\sum_{j>0,{\rm odd}}t_jz^{j}
\exp (-2\sum_{j>0,{\rm odd}}{\partial\over\partial t_j}{z^{-{j}}\over j})
\tau\\
&\quad
\otimes
\exp -\sum_{j>0,{\rm odd}}t_jz^{j}
\exp (2\sum_{j>0,{\rm odd}}{\partial\over\partial t_j}{z^{-{j}}\over j})
\tau=\tau\otimes\tau\, .
\end{aligned}
\end{equation}
Equation (\ref{BKPB}) is called the BKP  hierarchy in the bosonic picture.
It is straightforward, using change of variables and Taylor's
formula, to rewrite (\ref{BKPB}) into a generating series of Hirota bilinear equations on the tau-function
(see e.g. \cite{DJKM3}). However, we will not do that here, but rather concentrate on obtaining polynomial tau-functions of this hierarchy.

Define $H(t)=\sum_{j>0,{\rm odd}}\frac{t_j}2 \alpha_j$. 
Since
\[
e^{\sum_{j>0,{\rm odd}}\frac{t_j}2 \alpha_j} e^{-\sum\limits_{k<0, {\rm odd}}
\frac{\alpha_k}{k} z^{-k} }=e^{\sum_{j>0,{\rm odd}}t_jz^{j}}e^{-\sum\limits_{k<0, {\rm odd}}
\frac{\alpha_k}{k} z^{-k} }e^{\sum_{j>0,{\rm odd}}\frac{t_j}2 \alpha_j}\, ,
\]
one finds, using  the expression for $\phi(z)$ of Theorem  \ref{t2.3}, that 
\begin{equation}
\label{need}
e^{H(t)}\phi(z)e^{-H(t)}=e^{\sum_{j>0,{\rm odd}}t_jz^{j}}\phi(z),\quad e^{H(t)}|0\rangle=|0\rangle\, .
\end{equation}
Moreover, 
\begin{equation}
\label{ttau}
\tau(t)= \sigma(g|0\rangle)=\langle 0|e^{H(t)} g|0\rangle\,  .
\end{equation}

Let us calculate the following elements (cf. (\ref{orbit}))
\[
\langle 0|e^{H(t)} v_1v_2\cdots v_k|0\rangle\, .
\]
Define the elementary Schur polynomials $s_i(t)$  by $e^{\sum_{j=1}^\infty t_j z^j}=\sum_{i=0}^\infty s_i(t)z^i$ and let  $\tilde t=(t_1,0,t_3,0,t_5,\ldots)$.
Let $v_j$ be of the form 
\[ v_j=\phi_{-\lambda_j}+\sum_{n<\lambda_j} a_{-nj}\phi_{-n},\ \mbox{with }
a_{-sj}=0\ \mbox{for } s<<0\, .
\]
Since the map $(s_1(t),s_2(t),\ldots, s_k(t)):\mathbb{C}^k\to \mathbb{C}^k$ is surjective (actually an isomorphism), one can find constants $c_j=(c_{1j},c_{2j},\ldots)$ such that 
\[
1+\sum_{ n<\lambda_j} a_{-nj}z^{\lambda_j-n}=\sum_{k=0}^\infty s_k(c_j)z^k = e^{\sum_{i=1}^\infty  c_{ij}z^i}\, .
\]
Using this and (\ref{need}),
we have 
\begin{equation}
\label{calc}
\begin{aligned}
e^{H(t)}v_je^{-H(t)}&=e^{H(t)}\left(
\phi_{-\lambda_j}+\sum_{n<\lambda_j} a_{-nj}\phi_{-n}\right)e^{-H(t)}\\
&= {\rm Res}_{z=0}\,e^{H(t)}\phi(z)e^{-H(t)}\left(z^{-\lambda_j-1} +\sum_{ n<\lambda_j} a_{-nj}z^{-n-1}\right)dz\\
&= {\rm Res}_{z=0}\,\phi(z)e^{\sum_{\ell, {\rm odd}} t_\ell z^{\ell}}\left(z^{-\lambda_j-1} +\sum_{ n<\lambda_j} a_{-nj}z^{-n-1}\right)dz\\
&= {\rm Res}_{z=0}\,\phi(z)e^{\sum_{\ell, {\rm odd}} t_\ell z^{\ell}}z^{-\lambda_j-1} e^{\sum_{i=1}^\infty c_{ij}z^i}dz\\
&= {\rm Res}_{z=0}\,\phi(z)e^{\sum_{\ell, {\rm odd}} t_\ell z^{\ell}+ \sum_{i=1}^\infty c_{ij}z^i}z^{-\lambda_j-1}dz\\
&= {\rm Res}_{z=0}\,\phi(z)e^{\sum_{i=1}^\infty (\tilde t_i + c_{ij})z^i}z^{-\lambda_j-1}dz\\
&= {\rm Res}_{z=0}\,\phi(z)\sum_{\ell=0}^\infty s_\ell(\tilde t+c_j)z^{\ell-\lambda_j-1}dz\\
&= \sum_{\ell=0}^\infty s_\ell(\tilde t+c_j)\phi_{\ell-\lambda_j}:=v_j(\tilde t+c_j)\, .
\end{aligned}
\end{equation}
Hence the BKP tau-function that corresponds to an element of the form (\ref{orbit}) is the vacuum expectation value
\begin{equation}
\label{orbit2}
\langle 0| v_1(\tilde t+c_1)v_2(\tilde t+c_2)\cdots v_{k}(\tilde t+c_{k})|0\rangle\, .
\end{equation}
If $k=2\ell$, 
  this expectation value  is equal to (cf. the Appendix of \cite{DJKM3} or \cite{You})
\begin{equation}
\label{orbit3}
\langle 0| v_1(\tilde t+c_1)v_2(\tilde t+c_2)\cdots v_{2\ell}(\tilde t+c_{2\ell})|0\rangle={\rm Pf}\,\left( \langle 0|v_i(\tilde t+c_i)v_j(\tilde t+c_j)|0\rangle\right)_{ij}.
\end{equation}
Here Pf stands for the Pfaffian of a $2\ell\times2\ell$ skewsymmetric matrix $A=(a_{ij})$:
\[
{\rm Pf}\, (A)=\sum_{\sigma}{\rm sgn}(\sigma)\prod_{i=1}^\ell a_{\sigma(2i-1),\sigma(2i)},
\]
where we take the sum over all permutations  $\sigma$  for which  $\sigma(2i-1)<\sigma(2i+1)$ and $\sigma(2i-1)<\sigma(2i)$ for all $1\le i\le \ell$.
Note that in the Pfaffian of (\ref{orbit3}) we only take products of vacuum expectation values $\langle 0|v_i(\tilde t+c_i)v_j(\tilde t+c_j)|0\rangle$ with $i<j$.
If $k$ is odd there is a problem, the vacuum expectation value is nonzero, but we cannot relate this to a Pfaffian, since the Pfaffian of an odd skewsymmetric matrix is $0$. We solve this problem as follows.
We can assume that $k$ is always even; if $k$ is odd, we add the element $v_{k+1}= \phi_0$, since this only changes the vacuum expectation value by a scalar factor:
\[
 \langle0|e^{H(t)} v_1v_2\cdots v_kv_{k+1}|0\rangle
=\langle0|e^{H(t)} v_1v_2\cdots v_k\phi_0|0\rangle=\frac1{\sqrt 2} 
\langle 0|e^{H(t)} v_1v_2\cdots v_k|0\rangle.
\] 
In this way we get a vacuum expectation value of an even number of $v_j$'s, which we can relate to a Pfaffian which is nonzero.
Hence, from now on we assume, without loss of generality, that we have an even number of $v_j$'s. This means that that such an element is related to a strict partition
$\lambda=(\lambda_1,\lambda_2,\ldots,\lambda_{2\ell})$ with an even number of parts, where the last part is allowed to be zero. We call such partition  an extended strict partitions. Clearly there is a bijection between the this set and the collection of strict partitions. 

Let 
\[
\chi_{\lambda_i,\lambda_j}(c_i,c_j)=
\langle 0|v_i( c_i)v_j(c_j)|0\rangle\, .
\]
We have:
\[
\begin{aligned}
\chi_{\lambda_i,\lambda_j}(c_i,c_j)
=&
\sum_{m=0}^\infty s_m(c_i)\sum_{\ell=0}^\infty s_\ell(c_j)
\langle 0|\phi_{m-\lambda_i}\phi_{\ell-\lambda_j}|0\rangle\\
=&\sum_{m=\lambda_i}^\infty s_m(c_i)\sum_{\ell=0}^{\lambda_j} s_\ell(c_j)
\langle 0|\phi_{m-\lambda_i}\phi_{\ell-\lambda_j}|0\rangle\\
=&\sum_{\ell=0}^{\lambda_j} s_{\lambda_i+\lambda_j-\ell}(c_i) s_\ell(c_j)
\langle 0|\phi_{\lambda_j-\ell}\phi_{\ell-\lambda_j}|0\rangle\ ,.
\end{aligned}\]
Hence
\begin{equation}
\label{chi}
\chi_{\lambda_i,\lambda_j}(c_i,c_j)
=
\frac{1}2 s_{\lambda_i}(c_i) s_{\lambda_j}(c_j)+\sum_{\ell=1}^{\lambda_j} (-1)^{\ell}s_{\lambda_i+\ell}(c_i) s_{\lambda_j-\ell}(c_j)\, .
\end{equation}
Above we assume that $\lambda_i>\lambda_j\ge 0$; in this case we let $\chi_{\lambda_j, \lambda_i}=-\chi_{\lambda_i, \lambda_j}$.   This function is zero  if $\lambda_j< 0$.

As a result of Proposition \ref{prop} and the above considerations, we obtain the following 
\begin{theorem}
All polynomial tau-functions of the BKP hierarchy are, up to a scalar multiple, of the form 
\begin{equation}
\label{Pf}
{\rm Pf}\,\left( \chi_{\lambda_i,\lambda_j}(\tilde t+c_i,\tilde t+c_j)
\right)_{1\le i,j\le 2n},
\end{equation}
where $\lambda=(\lambda_1,\lambda_2,\ldots,\lambda_{2n})$  is an extended  strict partition, i.e., $\lambda_1>\lambda_2>\cdots > \lambda_{2n}\ge 0$,
$\tilde t=(t_1,0,t_3,0,t_5,0,\ldots)$, $c_i=(c_{1i},c_{2i},c_{3i},\ldots)$  are constants, and $\chi_{\lambda_i,\lambda_j}$ is given by (\ref{chi}).
\end{theorem}
{\bf Proof.} 
According to Proposition \ref{prop}, an element in the group orbit $O$ corresponds to an element in a $C_\lambda$ for $\lambda$ a strict partition.
Thus, we have to determine the image of the elements appearing in such a $C_\lambda$ under  the map $\sigma$ which appears in the neutral boson-fermion correspondence. According to (\ref{ttau}) and the calculations (\ref{calc}), this is equal to (\ref{orbit2}). Since we always may assume that $k$ is even (for odd $k$ we add the element $v_{k+1}(\tilde t+c_{k+1})=\phi_0$), using (\ref{orbit3}) we can express this in the Pfaffian. We thus obtain the desired result.\\

 \hfill$\square$
\\ 
\
\\
If one puts all constants $c_i$ and $c_j$ in (\ref{Pf}) equal to $0$, one obtains the $Q$-Schur functions, giving   a result of You \cite{You}  that all $Q$-Schur functions are tau-functions of the BKP hierarchy.

\begin{example}
One of the polynomial tau-functions  corresponding to the partition  $\lambda_1>\lambda_2>\lambda_3>0$, corresponds to
\[
 v_{1}v_{2}v_{3}|0\rangle=\sqrt 2
 v_{1}v_{2}v_{3}\phi_0|0\rangle\, .
\]
Add $\lambda_4=0$, then 
\[
\begin{aligned}
\sigma&( v_{1}v_{2}v_{3}|0\rangle)=\sqrt 2\langle 0| e^{H(t)}
 v_{1}v_{2}v_{3}\phi_0|0\rangle\\
&=\sqrt 2\langle 0| 
\sum_i s_i(\tilde t+c_1)\phi_{i-\lambda_1}\sum_j s_j(\tilde t+c_2)\phi_{j-\lambda_2}\sum_\ell s_\ell(\tilde t+c_3)\phi_{\ell-\lambda_3}
\phi_0|0\rangle\\
&=\sqrt 2\,
{\rm Pf}\,\left(\chi_{\lambda_i,\lambda_j}(\tilde t+c_i,\tilde t+ c_j)\right)_{1\le i,j\le 4}\\
&=\sqrt 2(
 \chi_{\lambda_1,\lambda_2}(\tilde t+c_1,\tilde t+c_2)\chi_{\lambda_3,0}(\tilde t+c_3,\tilde t)
-\chi_{\lambda_1,\lambda_3}(\tilde t+c_1,\tilde t+c_3)\chi_{\lambda_2,0}(\tilde t+c_2,\tilde t)+\\
&\qquad\qquad\qquad +
\chi_{\lambda_2,\lambda_3}(\tilde t+c_2,\tilde t+c_3)\chi_{\lambda_1,0}(\tilde t+c_1,\tilde t))\\
&=\sqrt 2(
 \chi_{\lambda_1,\lambda_2}(\tilde t+c_1,\tilde t+c_2)s_{\lambda_3}(\tilde t+c_3)
-\chi_{\lambda_1,\lambda_3}(\tilde t+c_1,\tilde t+c_3)s_{\lambda_2}(\tilde t+c_2)+\\
&\qquad\qquad\qquad +
\chi_{\lambda_2,\lambda_3}(\tilde t+c_2,\tilde t+c_3)s_{\lambda_1}(\tilde t+c_1))\, ,
\end{aligned}
\]
since $\chi_{\lambda_i,0}(\tilde t+c_i,\tilde t)=s_{\lambda_i}(\tilde t+c_i)$.
\end{example}

\section {Vertex operators and the DKP hierarchy
in the bosonic picture}\
In the spirit of  \cite{You2}, we embed $\oi$ into $\di$ by viewing $\oi$ as the fixed points of the involution $\iota$ on $DC\ell$ that sends $\phi_{\pm\frac12}$ to $\phi_{\mp\frac12}$ and fixes all other $\phi$'s.
Then, 
\[
\oi=\{x\in\di|\, \iota(x)=x\}.
\]
Define as in the $B$-case
\[
\begin{aligned}
\phi (z) =& \sum\limits_{j\in \mathbb{Z}} \phi_j z^{-j}:= \frac{1}{\sqrt 2}(\phi_{-\frac12}+\phi_{\frac12})+ \sum_{i=1}^\infty  \left(\phi_{-i-\frac12} z^i +(-1)^i \phi_{i+\frac12}z^{-i})\right), \\
\alpha(z) =& \sum\limits_{k\in 2 \mathbb{Z}+1} \alpha_k
 z^{-k-1} = :
\phi(z)  \frac{\phi(-z)}{z}:\, .
\end{aligned}
\]
There is a slight difference with the $B$-case in the sense that $V_D=V_0 \oplus V_1$ and that 
\[
\sigma (V_D)=\mathbb{C}[\theta,t_1,t_3,\ldots], \ \mbox{ where }\theta\ \mbox{ ia a Grassmann variable},
\]
 i.e., $ \theta^2=0$ and  $\theta$ commutes with all  $t_i$.
Then,
\[
\sigma \phi(z) \sigma^{-1} =\frac{1}{\sqrt 2} \left(\theta+\frac{\partial}{\partial\theta}\right)e^{\xi (t,z)} e^{-\eta
(t,z)},
\qquad  \sigma \frac{\phi_{-\frac12}-\phi_{\frac12}}{\sqrt 2} \sigma^{-1} =\frac{\theta-\frac{\partial}{\partial\theta}}{\sqrt 2}\, .
\]
Hence, 
\[
S_D=S_B-\frac12 \left(\theta-\frac{\partial}{\partial\theta}\right)\otimes  \left(\theta-\frac{\partial}{\partial\theta}\right)
\]
Hence equation (\ref{1.9}) turns into (\ref{BKPB})  for $\tau=\tau_0(t)$ or $\tau=\tau_1(t) \theta$
and (\ref{modDKP}) turns into 
\begin{equation}
\label{modB}
\begin{aligned}
{\rm Res}_{z=0}{dz\over z}&
\exp\left( \sum_{j>0,{\rm odd}}t_jz^{j} \right)
\exp \left(-2\sum_{j>0,{\rm odd}}{\partial\over\partial t_j}{z^{-{j}}\over j}\right)
\tau_0\\
&\quad
\otimes
\exp \left(-\sum_{j>0,{\rm odd}}t_jz^{j}\right)
\exp \left(2\sum_{j>0,{\rm odd}}{\partial\over\partial t_j}{z^{-{j}}\over j}\right)
\tau_1=\tau_1\otimes\tau_0\, .
\end{aligned}
\end{equation}
We now want to determine 
$\tau_0(t)=\sigma (v_1v_2\cdots v_{2\ell}|0\rangle)$ and 
$\tau_1(t)\theta=\sigma (v_1v_2\cdots v_{2\ell+1}|0\rangle)$ where $v_i$ is given by (\ref{orbit1}).
As in the BKP case,   
\[
\tau_0(t)=\langle 0|e^{H(t)}v_1v_2\cdots v_{2\ell}|0\rangle, \qquad \tau_1(t)=\langle 0|\phi_{\frac12}e^{H(t)}v_1v_2\cdots v_{2\ell+1}|0\rangle=
\]
Now 
\[
\begin{aligned}
v_j&=
\phi_{-\lambda_j-\frac12}+\sum_{n<\lambda_j} a_{-n-\frac12,j}\phi_{-n-\frac12}\\
&=\phi_{-\lambda_j}+\sum_{\lambda_j>n\in\mathbb{Z}} b_{-n,j}\phi_{-n} + b_j(\phi_{-\frac12} -\phi_{\frac12})\\
&= {\rm Res}_{z=0}\,\phi(z)\left(z^{-\lambda_j-1} +\sum_{ n<\lambda_j} b_{-nj}z^{-n-1}\right)dz  + b_j(\phi_{-\frac12} -\phi_{\frac12})\, ,
\end{aligned}
\]
where for $k\ge 1$:
\[
b_{-k,j}=a_{-k-\frac12,j},\ b_{k,j}=(-1)^ka_{k+\frac12,j},\ b_{0,j}=\frac{1}{\sqrt 2} (a_{\frac12,j}+a_{-\frac12,j}), \ b_{j}=\frac{1}{2} (a_{-\frac12,j}-a_{\frac12,j})\, .
\]
Then similarly to the calculations in (\ref{calc}), we obtain
\begin{equation}
\label{calc2}
\begin{aligned}
e^{H(t)}v_je^{-H(t)}
&= {\rm Res}_{z=0}\,e^{H(t)}\phi(z)e^{-H(t)}\left(z^{-\lambda_j-1} +\sum_{ n<\lambda_j} b_{-nj}z^{-n-1}\right)dz  + b_j(\phi_{-\frac12} -\phi_{\frac12})\\
&= {\rm Res}_{z=0}\,\phi(z)e^{\sum_{\ell, {\rm odd}} t_\ell z^{\ell}} \left(z^{-\lambda_j-1} +\sum_{ n<\lambda_j} b_{-nj}z^{-n-1}\right)dz  +b_j(\phi_{-\frac12} -\phi_{\frac12})\\
&= \sum_{\ell=0}^\infty s_\ell(\tilde t+c_j)\phi_{\ell-\lambda_j}+b_{j}(\phi_{-\frac12} -\phi_{\frac12}):=v_j(\tilde t+c_j; b_j)\, .
\end{aligned}
\end{equation}
Define, in a similar way as for $G=B$ for $i<j$ hence $\lambda_i>\lambda_j$:
\[
\rho_{\lambda_i,\lambda_j}(c_i,c_j; b_i,b_j)=
\langle 0|v_i( c_i;b_i)v_j(c_j;b_j)|0\rangle\, .
\]
We have:
\[
\begin{aligned}
\rho_{\lambda_i,\lambda_j}&(c_i,c_j;b_i,b_j)\\
=&
\langle 0|\left(\sum_{m=0}^\infty s_m(c_i)\phi_{m-\lambda_i}+b_i(\phi_{-\frac12} -\phi_{\frac12})\right)
\left(\sum_{\ell=0}^\infty s_\ell(c_j)\phi_{\ell-\lambda_j}+b_j(\phi_{-\frac12} -\phi_{\frac12})\right)|0\rangle\\
=&
b_js_{\lambda_i}(c_i) \langle 0|\phi_0\phi_{-\frac12}|0\rangle
-b_is_{\lambda_j}(c_j) \langle 0|\phi_{\frac12}\phi_0|0\rangle
-b_ib_j\langle 0|\phi_{\frac12}\phi_{-\frac12}|0\rangle\\
&\qquad\qquad+\sum_{m=\lambda_i}^\infty s_m(c_i)\sum_{\ell=0}^{\lambda_j} s_\ell(c_j)
\langle 0|\phi_{m-\lambda_i}\phi_{\ell-\lambda_j}|0\rangle\\
=&\left(s_{\lambda_i}(c_i)-\sqrt 2 b_i\right)\left(s_{\lambda_j}(c_i)+\sqrt 2 b_j\right)\langle 0|\phi_{0}\phi_{0}|0\rangle\\
&\qquad\qquad+\sum_{\ell=0}^{\lambda_j-1} s_{\lambda_i+\lambda_j-\ell}(c_i) s_\ell(c_j)
\langle 0|\phi_{\lambda_j-\ell}\phi_{\ell-\lambda_j}|0\rangle\, .
\end{aligned}\]
Hence
\begin{equation}
\label{chiD}
\rho_{\lambda_i,\lambda_j}(c_i,c_j;b_i,b_j)
=
 \left(\frac{s_{\lambda_i}(c_i)}{\sqrt 2}-b_i\right)\left(\frac{ s_{\lambda_j}(c_j)}{\sqrt 2}+b_j\right)+\sum_{\ell=1}^{\lambda_j} (-1)^{\ell}s_{\lambda_i+\ell}(c_i) s_{\lambda_j-\ell}(c_j)\, .
\end{equation}
Define 
\[
\rho_{\lambda_j}(c_j;b_j)=\langle 0|\phi_{\frac12}v_j(c_j;b_j)|0\rangle\, ,
\]
then 
\begin{equation}
\label{chiDsimple}
\rho_{\lambda_j}(c_j;b_j)=s_{\lambda_j}(c_j)\langle 0|\phi_{\frac12}\phi_0|0\rangle+b_j\langle 0|\phi_{\frac12}\phi_{-\frac12}|0\rangle=
\frac{ s_{\lambda_j}(c_j)}{\sqrt 2}+b_j\,.
\end{equation}
Thus we proved the following theorem. 
\begin{theorem}
\label{T7}
Let $\tilde t=(t_1,0,t_3,0,t_5,0,\ldots)$ and $c_i=(c_{1i},c_{2i},c_{3i},\ldots)$ and $b_i$  be constants. Denote by $\tilde c_j=\tilde t+c_j= (t_1+c_{1j}, c_{2j},t_3+c_{3j},c_{4j},\cdots)$.\\
(a) All polynomial tau-functions in $V_0$ of the DKP hierarchy are, up to a scalar multiple, of the form 
\begin{equation}
\label{Pff}
\rho_\lambda(\tilde t;c_1, \ldots ,c_{2n}; b_1,\ldots b_{2n}):={\rm Pf}\,\left( \rho_{\lambda_i,\lambda_j}(\tilde c_i,\tilde c_j;b_i,b_j)
\right)_{1\le i,j\le 2n},
\end{equation}
where $\lambda=(\lambda_1,\lambda_2,\ldots,\lambda_{2n})$  is an extended  strict partition, i.e., $\lambda_1>\lambda_2>\cdots > \lambda_{2n}\ge 0$,
and $\rho_{\lambda_i,\lambda_j}$ is given by (\ref{chiD}).\\
(b)
  All polynomial tau-functions in $V_1$ of the DKP hierarchy are, up to a scalar multiple, of the form 
\begin{equation} 
\label{Pfff}
\rho_\lambda(\tilde t;c_1, \ldots ,c_{2n+1}; b_1,\ldots b_{2n+1}):=
{\rm Pf}\,(A),
\end{equation}
where   $A$ is  a $(2n+2)\times  (2n+2)$ skewsymmetric  matrix, whose $k$-th row is equal to
\[
\begin{aligned}
\left(
-\rho_{\lambda_{k}}(\tilde c_{k};b_{b}),
-\rho_{\lambda_1,\lambda_k}(\tilde c_1,\tilde c_k;b_1,b_k),\cdots ,
-\rho_{\lambda_{k-1},\lambda_k}(\tilde c_{k-1},\tilde c_k;b_{k-1},b_k),
0, 
\right.\qquad&\\
\left. 
\rho_{\lambda_{k},\lambda_{k+1}}(\tilde c_{k},\tilde c_{k+1};b_{k},b_{k+1}),\cdots,
 \rho_{\lambda_{k},\lambda_{2n+1}}(\tilde c_{k},\tilde c_{2n+1};b_{k},b_{2n+1})
\right)&
\end{aligned}
\]
Again $\lambda=(\lambda_1,\lambda_2,\ldots,\lambda_{2n+1})$  is an extended  strict partition, i.e., $\lambda_1>\lambda_2>\cdots > \lambda_{2n+1}\ge 0$,
$\tilde t=(t_1,0,t_3,0,t_5,0,\ldots)$, $c_i=(c_{1i},c_{2i},c_{3i},\ldots)$ and $b_i$  are constants, and $\rho_{\lambda_i,\lambda_j}$ is given by (\ref{chiD}) and $\rho_{\lambda_i}$ is given by (\ref{chiDsimple}).\\
(c) Let $a_0,a_1\in\mathbb{C}$.  All polynomial tau functions of the MDKP hierarchy are pairs $(\tau_0,\tau_1)$, with either
\begin{equation}
\begin{aligned}
\tau_0&=a_0 \rho_{(\lambda_1, \ldots\lambda_{j-1},\lambda_{j+1},\ldots, \lambda_{2n+1})}(\tilde t;c_1, \ldots ,c_{j-1}, c_{j+1},\dots, c_{2n+1}; b_1, \ldots ,b_{j-1}, b_{j+1},\ldots, b_{2n+1}),\\
\tau_1&=a_1 \rho_{\lambda_1,\ldots,\lambda_{2n+1})}(\tilde t;c_1, \ldots ,c_{2n+1}; b_1,\ldots b_{2n+1}),
\end{aligned}
\end{equation}
or
\begin{equation}
\begin{aligned}
\tau_0&=a_0 \rho_{\lambda_1,\ldots,\lambda_{2n})}(\tilde t;c_1, \ldots ,c_{2n}; b_1,\ldots, b_{2n}),\\
\tau_1&=a_1 \rho_{(\lambda_1, \ldots\lambda_{j-1},\lambda_{j+1},\ldots, \lambda_{2n})}(\tilde t;c_1, \ldots ,c_{j-1}, c_{j+1},\dots, c_{2n}; b_1, \ldots ,b_{j-1}, b_{j+1},\ldots, b_{2n}).\\
\end{aligned}
\end{equation}
\end{theorem}
Theorem \ref{T7} generalizes the results of Y. You \cite{You2}, that $Q$-Schur functions are polynomial tau-functions of the DKP and MDKP hierarchies.

\end{document}